

\magnification=\magstep1
\overfullrule=0pt
\def\frac#1#2{{#1\over#2}}

\line{\hfill GSI-94-65}
\line{\hfill HD-TVP-94-19}

\vglue 1.0in
\centerline{{\bf $\pi-\pi$ scattering lengths at finite temperature}
\footnote{$^*$}{Supported in part by the Federal Ministry for Research
and Technology (BMFT), grant number 06 HD 742, and the Deutsche
Forschungsgemeinschaft, grant number Hu 233/4-3.}}

\medskip
\centerline{E. Quack\footnote{$^1$}{Present address: Gesellschaft f\"ur
Schwerionenforschung (GSI), P.O. Box 110552, D-64220 Darmstadt, Germany},
 P. Zhuang
\footnote{$^2$}{On leave from Huazhong Normal University, Wuhan, China;
Alexander von Humboldt fellow.   Present address: Institut f\"ur Theoretische
Physik, Universit\"at Regensburg, D-93040 Regensburg, Germany.},
 Y. Kalinovsky\footnote{$^3$}{Present address: Sektion Physik der
Universit\"at, Universit\"atsplatz 1,
D-18051 Rostock, Germany},
 S.P. Klevansky and J.H\"ufner,}

\smallskip
\centerline{Institut f\"ur Theoretische Physik, Philosophenweg 19,
D-69120 Heidelberg, Germany}

\vskip 0.5in
ABSTRACT:  The $s$-wave $\pi-\pi$ scattering lengths $a^I(T)$ at finite
temperature $T$ and isospin $I=0,2$ are calculated within the SU(2)
Nambu--Jona-Lasinio  model.
$a^2(T)$ displays a singularity at the Mott temperature $T_M$, defined as
$m_\pi(T_M) = 2m_q(T_M)$, while $a^0(T)$ is singular in addition at the lower
temperature $T_d$,
 where $m_\sigma(T_d) = 2m_\pi(T_d)$, $m_\sigma$ and $m_\pi$ being the
masses of the $\sigma$ and $\pi$ mesons, respectively.
Numerically we find $T_d = 198$MeV and $T_M=215$MeV.
We speculate on possible experimental consequences.

\baselineskip 20pt plus 4pt minus 4pt.
\eject

The scattering of pion by pions, $\pi + \pi\rightarrow
\pi+\pi$ is one of the most fundamental hadronic processes of
QCD at a mesonic level.  This is so for several reasons.  In particular, this
scattering involves only the lightest pseudoscalar modes of the theory, that
occupy a special role as being the Goldstone particles associated with the
almost exact $SU_L(2)\times SU_R(2)$ symmetry that is observed in the
particle spectrum.  As such, $\pi-\pi$ reactions provide a direct link between
the theoretical formalism of chiral symmetry and experiment.  This is
exemplified in the many existing studies of $\pi-\pi$ scattering using
chiral Lagrangians [1-7] that endeavour to calculate scattering amplitudes
and scattering lengths.  References [1-3] introduce [1] and use [2,3]
chiral perturbation theory (ChPT), while [4-6] calculate scattering lengths
in the Nambu--Jona-Lasinio (NJL) model [8,9].   Both $s$ and $p$ wave
scattering lengths are well reproduced in these models.  Essentially, the
Weinberg limit [10] can be recovered in all cases, and corrections to it,
due to processes such as pion rescattering are accounted for, for example
in [2].

To date, investigations performed have all been at zero temperature.  Our
interest here, however, is to examine $\pi-\pi$ scattering at {\it finite}
temperature, but at zero baryon density, with a view to describing the
baryon free central rapidity region occurring in ultra relativistic heavy ion
collisions, in
which pions are copiously formed.  In this calculation, we focus in particular
on those temperatures in the vicinity of the phase transition that takes the
system from a chirally
symmetric state  at high temperature to a phase in which
chiral symmetry is broken.  In our study, we will restrict ourselves to
the SU(2) version of the NJL model,
$$
{\cal L}(x) = \bar\psi(x)(i\not\partial - m_0)\psi(x) +
G[(\bar\psi(x)\psi(x))^2
+(\bar\psi(x) i\gamma_5\vec\tau\psi(x))^2],
\eqno(1)
$$
where G is a coupling strength of dimension [Mass]$^{-2}$, and $m_0$ is the
current quark mass.  Once again, we recall that although this model lacks
confinement and is non-renormalizable, it gives a good and transparent
description of the low energy meson sector.  One simply regulates
divergent integrals with a cutoff.  In our case, we will do so by restricting
the 3-momentum of a quark or antiquark, $|\vec p|<\Lambda$.  At a finite
temperature $T_c$, this model displays a second order phase transition
in the chiral limit  $(m_0=0)$ from a broken to a chirally symmetric state.
   At finite values of the current quark mass
$m_0$, the phase transition is washed out [11].
The temperature range around $T_c$, i.e. $150 - 200$ MeV, is of particular
importance for two reasons: (i) current experiments at CERN and AGS energies
appear to produce a hadron plasma in this temperature range, and (ii) in the
neighbourhood of second order phase transitions, divergences in measurable
quantities often occur, such as the phenomenon of critical opalescence [12].
In our case, we might expect that certain cross sections display a rapid
temperature dependence.   It is therefore of interest to us to study the
properties of $\pi-\pi$ scattering in the vicinity of this phase transition.
Our calculation of $\pi-\pi$ scattering is performed by keeping only the
diagrams from the lowest order expansion
in $1/N_c$, with $N_c$ the number of colors [13].

To establish our notation, we briefly summarize the basic formalism of
pion-pion scattering.  The reader may refer to Refs.[14]-[16] for a more
complete discussion.  Since three isospin channels are available for the
process $\pi\pi\rightarrow\pi\pi$, the invariant scattering amplitude
can be written in terms of three unknown functions,
$$
<c p_c; d p_d|{\cal T}|a p_a; b p_b> = {\cal T}_{ab;cd} =
A(s,t,u)\delta_{ab}\delta_{cd} + B(s,t,u)\delta_{ac}
\delta_{bd} + C(s,t,u)\delta_{ad}\delta_{bc}.
\eqno(2)
$$
Here $a,b$ and $c,d$ are the isospin labels of the initial and final states
respectively, and $s,t$ and $u$ are the usual Mandelstam variables,
$s=(p_a+p_b)^2$, $t=(p_a-p_c)^2$ and $u=(p_a-p_d)^2$.  Perfect crossing
symmetry enables one to express ${\cal T}_{ab;cd}$ in terms of one amplitude
alone,
since under the exchange $s\leftrightarrow t$, $A(s,t,u)=B(t,s,u)$, and
under $s\leftrightarrow u$, $A(u,t,s) = C(s,t,u)$.
It is a standard exercise to project out amplitudes of definite total
isospin $I$, which we denote as ${\cal T}^I(s,t,u)$.
In the limit of scattering at threshold, $\sqrt{s} = 2m_\pi$, $t=u=0$, the
${\cal T}^I$  approach the scattering lengths $a^I$, and one finds the
relation
$$
a^I = \frac 1{32\pi}{\cal T}^I(s=4m_\pi^2,t=0,u=0).
\eqno(3)
$$
To lowest order in $1/N_c$, the invariant amplitude ${\cal T}_{ab;cd}$ is
calculated
 from the box and $\sigma$-propagation diagrams shown in Fig.1.
Since these need only be evaluated at threshold, we have $p_i^2=p^2=m_\pi^2
=s/4$.  This choice of the pion momentum does not restrict generality at
$T=0$ since the system is Lorentz invariant.  For $T\ne0$, the heat bath
defines
a special system and the specification $p_i^2=p^2=s/4$ restricts our results to
the scattering of pions whose c.m. system is at rest in the heat bath.

{\it Box diagram}:  The box diagram of Fig.1 is generic for all diagrams
which can be constructed with internal quark and antiquark lines.  From
crossing relations, there
are three such possibilities, which we label 1,2 or 3,
and which correspond to the diagram shown, and those generated by
$s\leftrightarrow t$ and $s\leftrightarrow u$ exchanges, respectively.
Here, we follow the notation and calculation of Ref.[6].
  A direct evaluation of the diagrams leads one to the result
$$\eqalign{
({\cal T}_1)_{ab;cd} &=
(\delta_{ab}\delta_{cd}+\delta_{ac}\delta_{bd}-\delta_{ad}\delta_
{bc})[4N_cN_fig_{\pi q q}^4][I(0) + I(p)- p^2K(p)  ]\cr
({\cal T}_2)_{ab;cd} &=
(\delta_{ab}\delta_{cd}-\delta_{ac}\delta_{bd}-\delta_{ad}
\delta_{bc})[4N_cN_fig_{\pi q q}^4][ I(0) + I(p)-p^2K(p) ] \cr
({\cal T}_3)_{ab;cd} &=(-\delta_{ab}\delta_{cd} + \delta_{ac}\delta_{bd} +
\delta_{ad}\delta_{bc})[8N_cN_fig_{\pi q q}^4[I(0)+p^4L(p)/2-2p^2K(p)]\cr
}
\eqno(4)
$$
in terms of the integrals $I(p)$, $K(p)$ and $L(p)$,
$$\eqalign{
I(p) &= \int\frac{d^4k}{(2\pi)^4}\frac 1{[k^2-m^2][(k+p)^2-m^2]}\cr
K(p) &= \int\frac{d^4k}{(2\pi)^4}\frac 1{[k^2-m^2]^2[(k+p)^2-m^2]}\cr
L(p) &= \int\frac{d^4k}{(2\pi)^4}\frac 1{[k^2-m^2]^2[(k+p)^2-m^2]^2}.\cr
}
\eqno(5)
$$
For $T=0$, the integrals in Eq.(5) have been evaluated by [6] to obtain the
scattering lengths.
At finite temperatures, however, one has $\int{d^4k}/(2\pi)^4\rightarrow
\frac i\beta\sum_n\int d^3 k/(2\pi)^3$.  Here $\beta$ is the inverse
temperature, and  the sum on $n$ runs over the
Matsubara fermion frequencies $i\omega_n$, $\omega_n=(2n+1)\pi/\beta$,
$n=0,\pm1,
\pm2\cdots$, that occur in the 4-vector $k=(i\omega_n,\vec k)$.  Simple
analytic expressions can be obtained for these functions, and they are
listed in the appendix.

{\it $\sigma$-propagation  diagram:}  The  diagram
with intermediate $\sigma$ propagation shown in Fig. 1 can be
simply expressed in terms of its component parts.
  Again two further diagrams are generated by $s\leftrightarrow t$ and
$s\leftrightarrow u$ exchange. Defining the $\sigma-\pi-\pi$
vertex as
$$
\Gamma^{\sigma \pi \pi}(p',p) = N_cN_f\int\frac{d^4k}{(2\pi)^4} Tr
[S(k+p)\gamma_5S(k)\gamma_5S(k+p')],
\eqno(6)
$$
one may easily verify that the required choices of momenta for $s$ and
$t$ channel graphs lead to
$$\eqalign{
\Gamma^{\sigma \pi \pi}(p,-p) &= -8N_cN_f mI(p) \quad (s-\rm{channel})\cr
 &\rm{and} \cr
\Gamma^{\sigma\pi\pi}(p,p) &= -8N_cN_f m[I(0) - p^2K(p)] \quad
(t-\rm{channel}).\cr }
\eqno(7)
$$
Required also are the meson scattering amplitudes $D_\pi$ and
$D_\sigma$, defined as
$$
-iD_M(k) = \frac{2iG}{1-2G\Pi_M(k)},
\eqno(8)
$$
for $M=\sigma, \pi$.   Since it can be shown that $1-2G\Pi_M(k) = m_0/m
+4iGN_cN_f(k^2-\epsilon_M^2)I(k)$, with $\epsilon_\pi = 0$, $\epsilon_\sigma
= 4m^2$, and also that $m_\pi^2 = -m_0/m4iGN_cN_fI(m_\pi)$, one may express
the functions in (8) in terms of $I(p)$ only.  One has
$$
[-iD_M(k)]^{-1} = 2N_cN_f[(k^2-\epsilon_M^2)I(k) - m_\pi^2I(m_\pi)].
\eqno(9)
$$
Now one may construct the amplitudes for the scattering diagrams.  One has
$$
({\cal T}_4)_{ab;cd} = \delta_{ab}\delta_{cd} g_{\pi q q}^4 [\Gamma^{\sigma
\pi\pi}
(p,-p)]^2 D_\sigma(2p)
\eqno(10)
$$
in the $s$-channel, and $s\leftrightarrow t$ and $s\leftrightarrow u$-exchanges
give rise to
$$
({\cal T}_5)_{ab;cd} =(\delta_{ac}\delta_{bd} + \delta_{ad}\delta_{bc}) g_{\pi
q q}^4
[\Gamma^{\sigma\pi\pi}(p,p)]^2D_\sigma(0).
\eqno(11)
$$
Finally, expressions (4), (10) and (11) require the temperature dependent
pion-quark coupling $g_{\pi q q}$, and which, using the same arguments that
led to Eq.(9), is given as
$
g_{\pi q q}^{-4} = - N^2[I(0) + I(p) - m_\pi^2K(p)]^2.
$

It is now a simple matter to extract the functions $A(s,t,u)$, $B(s,t,u)$
and $C(s,t,u)$ by summing the amplitudes $({\cal T}_i)_{ab;cd}$ of (4), (10)
and
(11), and identifying $A$,$B$, and $C$ as the coefficients of
$\delta_{ab}\delta_{cd}$,
$\delta_{ac}\delta_{bd}$ and $\delta_{ad}\delta_{bc}$ respectively.  From
these, one can construct the ${\cal T}^I$, and via Eq.(3), the $s$-wave
scattering amplitudes $a^I$ can be
obtained.  One finds
$$\eqalign{
{\cal T}^0 &= 6{\cal T}_1 - {\cal T}_3 + 3{\cal T}_4 + 2{\cal T}_5 \cr
{\cal T}^1  &=0 \cr
{\cal T}^2 &= 2{\cal T}_3 + 2{\cal T}_5 \cr }
\eqno(12)
$$
where the ${\cal T}_i$ are the functions in Eqs.(4), (10) and (11), here to be
understood to be denuded of their isospin cofactors.   One observes that ${\cal
T}^1$ is identically zero, as it must be for $s$-waves.

Individual contributions to the scattering amplitude  are shown as a function
of temperature in Fig.2.  For this calculation, a standard set of parameters
has
been used, viz. $\Lambda=631$MeV, $m_0=5.5$MeV and $G=5.514$GeV$^{-2}$ that
lead
to a value of the current quark mass $m=339$MeV, $m_\pi=139$MeV and
$f_\pi=93.3$MeV.
For $m_0=0$, these values of $\Lambda$ and $G$ lead to a phase transition
at $T_c=195$MeV.  One observes that the ${\cal T}$-matrix amplitudes are
constant at low temperatures, and are approximately equal in magnitude at
$T=0$.
Structure appears in the amplitudes ${\cal T}_i$ at high values of the
temperature.  ${\cal T}_1$ and ${\cal T}_3$ are negative and divergent at the
Mott temperature $T_M$, defined as the temperature at which the pion mode
enters into the continuum, thereby dissociating into constituent quark and
antiquark, i.e.
$$
m_\pi(T_M) = 2 m_q(T_M).
\eqno (13)
$$
The amplitude ${\cal T}_5$ is positive and divergent at this temperature.
On the other hand ${\cal T}_4$ diverges at the temperature $T_d$, at which
the $\sigma$-meson can dissociate into pions,
$$
m_\sigma(T_d) = 2 m_\pi(T_d),
\eqno(14)
$$
due to the $s$-channel pole that occurs at this point, see Eq.(11).  For our
set of parameters, we find $T_d = 198$MeV, while $T_M=215$MeV.

In Fig.3, we plot the scattering lengths $a^{I=0,2}$ as a function of the
temperature.
At $T=0$, we obtain the numerical values
$$\eqalign{
a^0&=0.179 \cr
&{\rm  and } \cr
a^2&=-0.047,  \cr}
\eqno(15)
$$
 in units
of the pion mass.  These are close to the Weinberg values
$(a^0)^W=7m_\pi^2/(32\pi
f_\pi^2)=0.16$ and $(a^2)^W=-2m_\pi^2/(32\pi f_\pi^2)=0.044$, as one would
expect [4,6], with the
difference being attributable to the composite structure of the mesons.
Experimentally, one has
[17] $a^0=
0.26\pm0.05$ and $a^2 = - 0.028\pm0.012$,
which are compatible with those from a later reference [18],  $a^0=0.20\pm0.01$
and
$a^2=-0.037\pm0.004$.  By comparison, chiral perturbation
theory for SU(2) [3] gives $a^0 = 0.20$, and $a^2 = -0.042$.
  It is seen that our results underestimate
the ChPT and experimental results.   This may be so since our calculation
contains only the lowest order terms in $1/N_c$, and  does not include any
higher order physical processes, such as pion rescattering [2].   We
nevertheless
 have the advantage over ChPT in that we are able to calculate our model
results over the entire temperature range.

Regarding the temperature dependence of the $a^I$, we see firstly that $a^0$
varies only slowly with temperature until $T\simeq 140$MeV, and then displays a
steep singularity at $T=T_d$, due to the fact that this amplitude contains
${\cal T}_4$, or physically, that it contains the $s$-channel pole due
to $\sigma$ exchange.  It diverges yet again at the Mott temperature $T_M$,
since it also contains the amplitudes $T_1$, $T_3$ and $T_5$.  On the other
hand, $a^2$ is seen to vary slowly with temperature for temperatures less than
$T\simeq 200$MeV.  Since this scattering length also contains the amplitudes
$T_3$ and $T_5$, it  exhibits a divergence at the Mott temperature.

  We  compare our calculation with the zeroth order
forms for $(a^I)^W$ given earlier in Eq.(15), by blindly  extrapolating them to
finite
temperature
with the replacement $m_\pi\rightarrow m_\pi(T)$ and $f_\pi\rightarrow
f_\pi(T)$, as calculated in the NJL model.
A direct calculation of $f_\pi$ at finite temperature [19] indicates that
$f_\pi$ is monotonically decreasing with temperature, and diverges at the Mott
temperature.  Thus one sees that the extrapolated Weinberg estimate follows
the calculation for $a^2$ rather closely.  On the other hand, it is unable to
reproduce the singular structure at $T_d$ for $a^0$.
 One should also note that
  this naive extension of the Weinberg formula to finite temperature
makes a definite prediction, i.e. that the ratio $a^0(T)/a^2(T)=-3.5$ is a
constant, independent of temperature.  This appears to hold at the
$10\%$ level for temperatures up to $T\simeq 145$ MeV, and is thus an excellent
 approximation at low temperatures.  It breaks down of course in the
vicinity of $T_d$ for $a^0$.
  We comment that similar trends are observed at
finite densities, but at $T=0$ [5], where a phase transition also occurs.

The results shown in Fig.3 have been obtained in a particular model, the NJL
model, which  has proven rather reliable at the temperature $T=0$.  One may
ask the question to what degree the results in the neighbourhood of the
phase transition can be trusted, in particular in view of the fact that the
NJL model only describes the chiral aspect of the phase transition but not
deconfinement.   The feature that the scattering lengths diverge
at the Mott temperature may have a simple geometrical origin.  Physically,
when the pion reaches threshold, its constituents become unbound, and the
pion radius becomes infinite.  This is expressed in the NJL model via the
relation of the charge radius to $f_\pi$, i.e. $f_\pi^2 \propto 1/<r^2>$, where
$<r^2>$ is the mean pion charge radius squared.  Using the Weinberg
extrapolation, which according to our calculation is a good (but not perfect)
approximation of our calculated results, we have that the $|a(T)|$ increase
like
$|a(T)|\propto <r^2>(T)$, with $a=a^0$ or $a^2$.  This increase of $|a(T)|$ in
the neighbourhood of the
chiral transition is thus really a consequence of the pion becoming unbound,
i.e. it incorporates implicitly the signal of deconfinement.  Therefore the
increase of $|a(T)|$, $T\rightarrow T_M$ has a rather clearly understood
physical
origin, and may thus survive any model dependence.  However, the {\it detailed}
structure of singularities as shown on Fig.3 may reflect some particular
features of the NJL model.

We conclude this paper by speculating as to whether there is a possibility of
experimentally verifying the predicted singularities of the $\pi-\pi$
scattering
lengths in the neighbourhood of the chiral (and deconfinement) phase
transition.
The signal observed in the Hanbury-Brown-Twiss (HBT) experiment of $\pi-\pi$
correlations is related to (a) the space-time structure of the pion emitting
system, and to (b) the pion-pion final state interaction.  Any anomaly observed
in the HBT results may be a hint as to the importance of (b), or indirectly
the $\pi-\pi$ scattering lengths.  One may perhaps even try to compare events
with
different pion temperatures.

\vskip 0.1in

\noindent
Acknowledgments:

\vskip 0.2in
\noindent
J.H. and S.P.K. are grateful to M. Volkov for interesting and illuminating
discussions.  One of us (Y.K.) thanks the Max-Planck-Gesellschaft for its
financial support during the course of this work.

\vskip 0.2in
\noindent
Appendix

The temperature dependent functions $I(p)$, $K(p)$ and $L(p)$ are listed
here for the kinematics at threshold, $p=(p_0,\vec p) = (\sqrt s/2,0)$.  One
finds
$$
-iI(p_0) = \int\frac{dk}{2\pi^2} \frac {k^2}{E}\frac
{f(E) - f(-E)}{p_0^2-4E^2},
\eqno(A1)
$$
$$\eqalign{
-iK(p_0) = -\int \frac{dk}{2\pi^2}&\frac{k^2}{4E^3}[f(E) - f(-E)]
[\frac 1{p_0^2-4E^2} - \frac{8E^2}{(p_0^2-4E^2)^2}  \cr
 &+ 2\beta Ef(E)f(-E)
\frac1{p_0^2-4E^2}],\cr }
\eqno (A2)
$$
and
$$\eqalign{
-iL(p_0) = &-\int\frac{dk k^2}{4\pi^2}\frac 1{4E^3p_0^2}[(f(E)-f(-E))
[\frac 1{p_0^2 - 4E^2} -\frac{12E^2}{(p_0^2-4E^2)^2} - \frac{64E^4}{(p_0^2
-4E^2)^3}] \cr &+ \beta Ef(E)f(-E)(\frac1{p_0^2-4E^2} +
\frac{8E^2}{(p_0^2-4E^2)^2})] \cr }
\eqno(A3)
$$
In these expressions, $f$ is the Fermi function, $f(E) = [\exp(\beta E)-1]^
{-1}$.

\vskip 0.2in

References.
\vskip 0.2in

\noindent
[1] J. Gasser and H. Leutwyler, Ann. Phys. (N.Y.) 158 (1984) 142\hfill\break
[2] J.F. Donoghue, C. Ramirez and G. Valencia, Phys. Rev. {\bf D38} (1988)
2195 \hfill\break
[3] V. Bernard, N. Kaiser and U.-G. Meissner, Nucl. Phys. {\bf B364} (1991)
283\hfill\break
[4] V. Bernard, U.-G. Meissner, A.H. Blin and B. Hiller, Phys. Lett. {\bf
B253} (1991) 443\hfill\break
[5] M.C. Ruivo, C.A. de  Sousa, B. Hiller and A.H. Blin, Nucl. Phys. {\bf
A575} (1994) 460\hfill\break
[6] H.J. Schulze, Pion-pion scattering lengths in the Nambu-Jona-Lasinio
model, Preprint, 1994.\hfill\break
[7] C.D. Roberts, R.T. Cahill, M.E. Sevior and  N. Iannella, Phys. Rev.
{\bf D49} (1994) 125\hfill\break
[8] Y. Nambu and G. Jona-Lasinio, Phys. Rev. {\bf 122} (1961) 345;{\it ibid}
{\bf 124} (1961) 246  \hfill\break
[9] For a review, see U. Vogl and W. Weise, Prog. Part. Nucl. Phys.
{\bf 27} (1991) 195, S.P. Klevansky, Rev. Mod. Phys. {\bf 64} (1992) 649; T.
Hatsuda and T. Kunihiro, Phys. Rep., to appear\hfill\break
[10] S. Weinberg, Phys. Rev. Lett. {\bf 17} (1966) 616\hfill\break
[11] P. Zhuang, J. H\"ufner and S.P. Klevansky, Nucl. Phys. A, in press
\hfill\break
[12] H. Stanley, {\it Introduction to phase transitions and critical
phenomena}, (Oxford University press, NY 1971)  \hfill\break
[13] E. Quack and S.P. Klevansky, Phys. Rev. {\bf C49} (1994) 3283 \hfill\break
[14] J. L. Petersen, Phys. Rep. {\bf C2} (1971) 155\hfill\break
[15] B.R. Martin, D. Morgan and G. Shaw, {\it Pion-pion interactions
in particle physics}, (Academic Press, London, 1976). \hfill\break
[16] H. Pagels, Phys. Rep. {\bf 16} (1975) 219\hfill\break
[17] C. Riggenbach, J. Gasser, J.F. Donoghue and B.R. Holstein, Phys.Rev.
{\bf D43} (1991) 127 \hfill\break
[18] M.E. Sevior, Nucl. Phys. {\bf 543}, (1992) 275c\hfill\break
[19] T. Hatsuda and T. Kunihiro, Phys. Lett. {\bf B185} (1987) 304. \hfill
\break

\vskip 0.2in
Figure Captions.
\vskip 0.2in

Fig.1:  Box and $\sigma$-propagation  diagrams for $\pi-\pi$ scattering.

\vskip 0.1in
Fig.2:  T matrix amplitudes in units of the pion mass, shown as a function of
temperature.  The Mott temperature $T_M$ and $\sigma$ dissociation temperature
$T_d$ are indicated
by the dashed vertical lines.

\vskip 0.1in
Fig.3:  Scattering amplitudes $a^0$ and $a^2$, in units of the pion
mass, shown as a function
of temperature.  The Mott temperature $T_M$ and $\sigma$ dissociation
temperature $T_d$ are indicated
by the vertical dashed lines.  Experimental values (see text) at $T=0$ are
also indicated.

\vfill
\bye